\documentclass[fleqn,usenatbib]{mnras}

\usepackage{newtxtext,newtxmath}
\usepackage[T1]{fontenc}
\usepackage{ae,aecompl}

\usepackage{graphicx}	% Including Figure~files
\usepackage{amsmath}	% Advanced maths commands
\usepackage{amssymb}	% Extra maths symbols
\usepackage{comment}

\title[(Non)existence of high-latitude open clusters]{The GALAH Survey and \textit{Gaia} DR2: (Non)existence of five sparse high-latitude open clusters}

\author[J. Kos et al.]{
Janez Kos,$^{1}$\thanks{E-mail: janez.kos@sydney.edu.au}
Gayandhi de Silva,$^{1,2}$
Joss Bland-Hawthorn,$^{1,3,4}$
Martin Asplund,$^{5,3}$\newauthor
Sven Buder,$^{6,7}$
Valentina D'Orazi,$^{8}$
Ly Duong,$^{5}$
Ken Freeman,$^{5}$
Geraint F. Lewis,$^{1}$\newauthor
Jane Lin,$^{5,3}$
Karin Lind,$^{6,9}$
Sarah L. Martell,$^{10,3}$
Katharine J. Schlesinger,$^{5}$\newauthor
Sanjib Sharma,$^{1}$
Jeffrey D. Simpson,$^{2}$
Daniel B. Zucker,$^{11}$
Toma\v{z} Zwitter,$^{12}$\newauthor
Timothy R. Bedding,$^{1,14}$
Klemen \v{C}otar,$^{12}$
Jonathan Horner,$^{13}$
Thomas Nordlander,$^{5,3}$\newauthor
Denis Stello,$^{10,14}$
Yuan-Sen Ting,$^{15,16,17}$
and Gregor Traven$^{12}$\\
\\
$^{1}$Sydney Institute for Astronomy, School of Physics A28, The University of Sydney, NSW 2006, Australia\\
$^{2}$Australian Astronomical Observatory, 105 Delhi Rd, North Ryde, NSW 2113, Australia\\
$^{3}$ARC Centre of Excellence for All Sky Astrophysics in 3 Dimensions (ASTRO-3D), Australia\\
$^{4}$Sydney Astrophotonic Instrumentation Labs, School of Physics, A28, The University of Sydney, NSW 2006, Australia\\
$^{5}$Research School of Astronomy \& Astrophysics, Australian National University, ACT 2611, Australia\\
$^{6}$Max Planck Institute for Astronomy (MPIA), Koenigstuhl 17, 69117 Heidelberg, Germany\\
$^{7}$Fellow of the International Max Planck Research School for Astronomy \& Cosmic Physics at the University of Heidelberg, Germany\\
$^{8}$Istituto Nazionale di Astrofisica, Osservatorio Astronomico di Padova, vicolo dell'Osservatorio 5, 35122, Padova, Italy\\
$^{9}$Department of Physics and Astronomy, Uppsala University, Box 516, SE-751 20 Uppsala, Sweden\\
$^{10}$School of Physics, UNSW, Sydney, NSW 2052, Australia\\
$^{11}$Department of Physics and Astronomy, Macquarie University, Sydney, NSW 2109, Australia\\
$^{12}$Faculty of Mathematics and Physics, University of Ljubljana, Jadranska 19, 1000 Ljubljana, Slovenia\\
$^{13}$University of Southern Queensland, Toowoomba, Queensland 4350, Australia\\
$^{14}$ Stellar Astrophysics Centre, Department of Physics and Astronomy, Aarhus University, Denmark\\
$^{15}$Institute for Advanced Study, Princeton, NJ 08540, USA\\
$^{16}$Department of Astrophysical Sciences, Princeton University, Princeton, NJ 08544, USA\\
$^{17}$Observatories of the Carnegie Institution of Washington, 813 Santa Barbara Street, Pasadena, CA 91101, USA\\
}

\date{Accepted XXX. Received YYY; in original form ZZZ}

\pubyear{2018}

\begin{document}
\label{firstpage}
\pagerange{\pageref{firstpage}--\pageref{lastpage}}
\maketitle

% Abstract of the paper
\begin{abstract}
Sparse open clusters can be found at high galactic latitudes where loosely populated clusters are more easily detected against the lower stellar background. Because most star formation takes place in the thin disk, the observed population of clusters far from the Galactic plane is hard to explain. We combined spectral parameters from the GALAH survey with the \textit{Gaia}~DR2 catalogue to study the dynamics and chemistry of five old sparse high-latitude clusters in more detail. We find that four of them (NGC~1252, NGC~6994, NGC~7772, NGC~7826) - originally classified in 1888 - are not clusters but are instead chance projections on the sky. Member stars quoted in the literature for these four clusters are unrelated in our multi-dimensional physical parameter space; the published cluster properties are therefore irrelevant. We confirm the existence of morphologically similar NGC~1901 for which we provide a probabilistic membership analysis. An overdensity in three spatial dimensions proves to be enough to reliably detect sparse clusters, but the whole 6-dimensional space must be used to identify members with high confidence, as demonstrated in the case of NGC~1901.
\end{abstract}

% Select between one and six entries from the list of approved keywords.
% Don't make up new ones.

%%%%%%%%%%%%%%%%%%%%%%%%%%%%%%%%%%%%%%%%%%%%%%%%%%

%%%%%%%%%%%%%%%%% BODY OF PAPER %%%%%%%%%%%%%%%%%%

\section{Introduction}

The GALAH survey \citep{desilva15,buder18} is a high-resolution (R=28,000), high signal-to-noise ratio (SNR$\approx$100) spectroscopic survey of one million stars. Its aim is to measure the abundances of up to 31 elements with a goal to disentangle the chemical history of the Milky Way \citep{freeman02}.  
Observations specifically targeting open clusters are carried out as part of a dedicated program (De Silva et al., 2018, in preparation) associated with the full GALAH survey.
Such clusters play a fundamental role in our understanding of the chemical evolution of stars, since they are almost the only stellar population with homogeneous elemental abundances \citep{desilva06,desilva07,sestito07,bovy16} arising from a common birth-time and place. Hence, processes in the evolution of stellar systems are best studied in clusters, including, but not limited to, the initial mass function of stars \citep{chabrier03, krumholz14}, the interaction with the disk \citep{gieles07, gieles08} or Galactic tidal fields \citep{baumgardt03}, the creation of blue stragglers \citep{stryker93}, initial binary fraction \citep{hurley07,fregeau09}, radial migration \citep{fujii12}, mass loss \citep{miglio17}, and atomic diffusion \citep{motta18}. Open clusters have long been considered as representatives of star formation in the Galaxy, because they are mostly found in the thin disk. Open clusters in other components of the Galaxy are rare, and so confirming their reality and measuring their properties is vital for  using them to study the aforementioned processes in parts of the Galaxy outside the Galactic plane.

\begin{figure}
\centering
\includegraphics[width=\columnwidth]{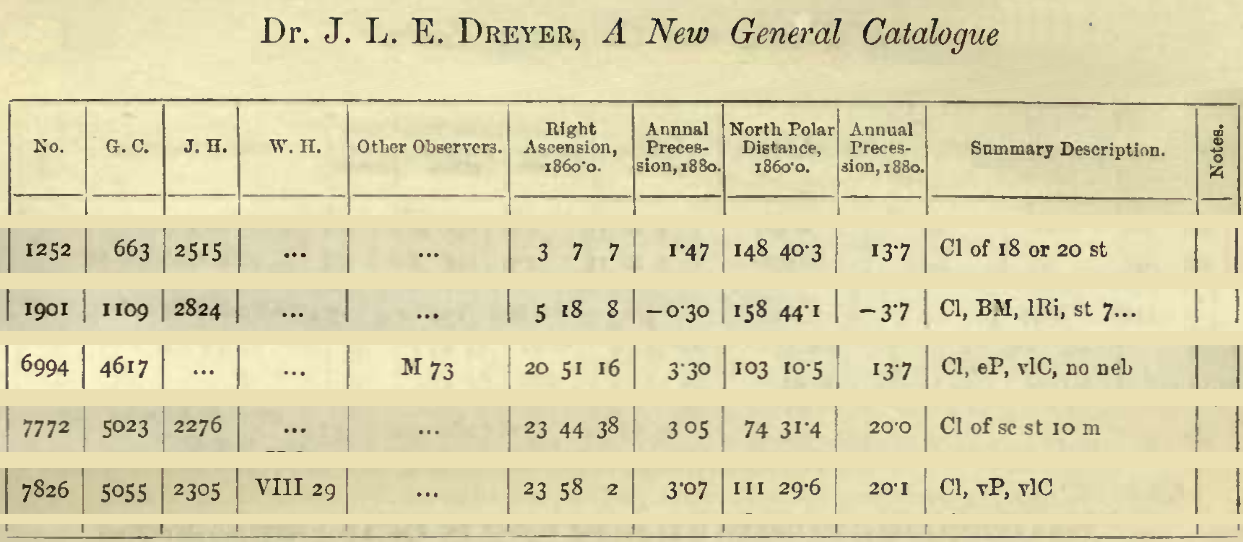}
\caption{For early observers these were without much doubt clusters of stars. The mistake could be attributed to ``[...] accidental errors occasionally met with in the observations of the two \textsc{Herschels}, and which naturally arose from the construction of their instruments and the haste with which the observations often necessarily were made.'' \citep{dreyer88}}
\label{fig:ngc}
\end{figure}

The clusters addressed in this work are generally believed to be old and far from the Galactic plane. At first glance, this is surprising because the survival time of such clusters is lower than their thin disk counterparts \citep{medina17}. A simple model where most star formation happens in the thin disk has to be updated to explain open clusters at high galactic latitudes, assuming these are real\footnote{Clusters can have high galactic latitude and still be well inside the thin disk if they are close to us. By high-latitude clusters, we mean those that are far from the Galactic plane.}. The prevailing theories are heating of the disk \citep{gustafsson16}, soft lifting through resonances \citep{medina16}, mergers \citep{keres05,sansici08}, and in-situ formation from high-latitude molecular clouds \citep{camargo16}. Sparse clusters are very suitable for testing the above theories because they are numerous and low in mass, and are therefore more likely to be involved in some of the above processes. 

Sparse clusters are the last stage at which we can currently connect the related stars together before they dissipate into the Galaxy. Chemical tagging -- the main objective of the GALAH survey -- often uses such structures as a final test before attempting to tag field stars. Because sparse clusters are are likely to be dissolving rapidly at the current epoch, we can expect to find many former members far from the cluster centre \citep{kos18,desilva11}. If these former members can be found, the dynamics of cluster dissipation and interaction with the Galactic potential could be studied in great detail.

Historically, sparse open clusters and open cluster remnants have been identified based on the aggregation of stars on the sky and their positions in the HR diagram. Spectroscopy has rarely been used, possibly because it was too expensive use of telescope time. Massive spectroscopic surveys either did not exist or only included one or a few potential members. Often, there was only one spectroscopically observed star in the cluster, so even for misidentified clusters the values for radial velocity ($v_r$) and metallicity\footnote{In the literature the metallicity $\left[ \mathrm{M} / \mathrm{H}\right]$ and iron abundance $\left[ \mathrm{Fe} / \mathrm{H}\right]$ are used interchangeably. We discuss the importance of differentiating between them in Section~\ref{sec:ngc1901members}.} $\left(\left[\mathrm{M} / \mathrm{H}\right], \left[ \mathrm{Fe} / \mathrm{H}\right]\right)$ can be found in the literature, even though the most basic spectroscopic inspection of a handful of stars would disprove the existence of the cluster. When identifying the clusters based on the position of stars in the HR diagram, most stars that fell close to the desired isochrone were used. Sparse clusters are often reported at high galactic latitudes, where the density of field stars is low and only a few are needed for an aggregation to stand out. Lacking other information, the phenomenon of pareidolia led early observers to conclude that many ``associations'' were real (Figure~\ref{fig:ngc}). The cluster labels have stuck, however, and several studies have been made of properties of these clusters (see Section~\ref{sec:clusters} and references therein). It has happened before, that after more careful investigation that included proper motions and radial velocities, a cluster has been disproved \citep[e.g. ][]{han16}. 

The amount of available data has increased enormously with \textit{Gaia} DR2 \citep{gaia18}. The search for new clusters will now be more reliable, given that precise proper motions and parallax add three more dimensions in which the membership can be established \citep{koposov17, castro18, torrealba18, gaudin18, dias18}. 

In Section~\ref{sec:e} we review the studied data and explain the methods used to disprove the existence of four clusters and confirm a fifth. Section~\ref{sec:ngc1901members} provides more details about the analysis of the real cluster NGC~1901. In Section~\ref{sec:d} we discuss some implications of our findings.

\section{Existence of clusters}
\label{sec:e}

\subsection{Data}

Stars observed as a part of the GALAH survey are selected from the 2MASS catalogue \citep{skrutskie06}. Depending on the observing mode, all the stars in a $1^\circ$ radius field are in a magnitude range $12<V<14$ for regular fields and $9<V<12$ for bright fields; for full details on the survey selection process, and the manner in which observations are taken, we direct the interested reader to \citet{martell17, buder18}. For some clusters targeted in the dedicated cluster program, we used a custom magnitude range to maximise the number of observed members. The $V$ magnitude is calculated from the 2MASS photometry. In general, the GALAH $V$ magnitude is $\sim0.3$ fainter than \textit{Gaia} $G$ magnitude.

Most of the data used in this work comes from the \textit{Gaia}~DR2 \citep{gaia18}, which includes positions, $G$ magnitudes, proper motions and parallaxes for more than 1.3 billion stars. This part of the catalogue is essentially complete for $12<G<17$, which is the range where we expect to find most of the cluster stars discussed in this paper. There are, however, a few members of the four alleged clusters that are brighter than $G=12$ and are not included in \textit{Gaia}~DR2, but do not impact the results of this paper. Radial velocities in \textit{Gaia}~DR2 are only given for 7.2 million stars down to $G=13$ (the limit depends on the temperature as well). Because the precision of radial velocities is significantly higher in the GALAH survey \citep{zwitter18} than the \textit{Gaia} data release, we use GALAH values wherever available. Because GALAH has a more limited magnitude range than \textit{Gaia}, there are many stars for which \textit{Gaia}~DR2 radial velocities must be used. From the cluster stars used in this work that have radial velocity measured in both \textit{Gaia} DR2 and GALAH we find no systematic differences larger than 0.2~$\mathrm{km\,s^{-1}}$ between the two surveys, so we can use whichever velocity is available.

\subsection{Clusters}
\label{sec:clusters}

In the fields observed by the GALAH survey, we identified 39 clusters with literature references. Four of them (NGC~1252, NGC~6994, NGC~7772, and NGC~7826) appeared to have no observed members -- stars clumped in the parameter space -- even though we targeted them based on the data in the literature. All four clusters are sparse (with possibly only $\sim10$ members), are extended, and at high galactic latitudes. Among the other 35 observed clusters we only found one (NGC~1901) that is morphologically similar to those four and it is most certainly a real open cluster, as confirmed by the literature \citep{eggen96, pavani01, dias02, carraro07, kharchenko13, conrad14} and our own observations. See Table~\ref{tab:clusters} for a list of basic parameters of these five clusters. We observed more clusters at high latitudes (Blanco~1, NGC~2632, M~67, and NGC~1817), but they are all more populated than the five clusters studied here.

\begin{table}
\setlength{\tabcolsep}{4.6pt}
\caption{Coordinates of the clusters used in this work. For the non-existing clusters the heliocentric distances $d$ and distances from the Galactic plane $z$ are taken from the literature (see Section~\ref{sec:clusters}).}
\centering
\small
\begin{tabular}{ccccccc}
\hline\hline
Cluster & $\alpha$ & $\delta$ & $l$ & $b$ & $d$ & z\\
 name & $^\circ$ & $^\circ$ & $^\circ$ & $^\circ$ & pc & pc\\
\hline
NGC~1252 & 47.704 & -57.767 & 274.084 & -50.831 & 1000 & -775 \\
NGC~1901 & 79.490 & -68.342 & 278.914 & -33.644 & 426.0 & -236.0\\
NGC~6994 & 314.750 & -12.633 & 35.725 & -33.954 & 620 & -345\\
NGC~7772 & 357.942 & 16.247 & 102.739 & -44.273 & 1500 & -1050\\
NGC~7826 & 1.321 & -20.692 & 61.875 & -77.653 & 600 & -590\\
\hline
\end{tabular}
\label{tab:clusters}
\end{table}

\begin{itemize}
\item NGC~1252 was thought to be metal poor, old (3 Gyr), and far from the Galactic plane ($z=-900$~pc). This would be a unique object, as no such old clusters are found that far from the plane \citep{delafuente13}. It has been argued in the past that this cluster is not real \citep{baumgardt03}.

\item NGC~6994 was thought to be old (2 to 3 Gyr), relatively far away from the plane ($z=-350$~pc) and a dynamically well evolved cluster remnant \citep{bassino00}. \citet{carraro00, odenkirchen02} successfully argue that NGC~6994 is neither a cluster or a remnant, but only an incidental overdensity of a handful of stars.

\item NGC~7772 was thought to be 1.5 Gyr old, more than 1 kpc below the plane and depleted of low mass stars \citep{carraro02}. 

\item NGC~7826 has never been studied in detail. It is included in some open cluster catalogues \citep{dias14}, in which an age of 2 Gyr and a distance from the Galactic plane of $z=-600$~pc are assumed.
\end{itemize}

\subsection{Literature members}

In Tables \ref{tab:members1252} to \ref{tab:members7826} we review the available memberships from the literature for clusters NGC~1252, NGC~6994, NGC~7772, and NGC~7826. It is clear they are not clusters and the possible members are in no way related, based on the \textit{Gaia} DR2 parameters and occasionally GALAH radial velocities. We cross matched the lists of members given in the literature to \textit{Gaia} DR2 targets based on positions and magnitude, so all the parameters given in Tables \ref{tab:members1252} to \ref{tab:members7826} are from \textit{Gaia} DR2 or GALAH. A small fraction of members in the literature can not be cross-matched with \textit{Gaia} DR2, either because the star is absent from \textit{Gaia} DR2, the coordinates in the literature are invalid, or the members are not clearly marked in a figure or table.

In contrast to our findings for the four other clusters, our results confirm that NGC~1901 is a real cluster, and provide our own membership analysis in Section~\ref{sec:ngc1901members}.

\subsection{6D parameter space}

\begin{figure*}
\centering
\includegraphics[width=0.92\textwidth]{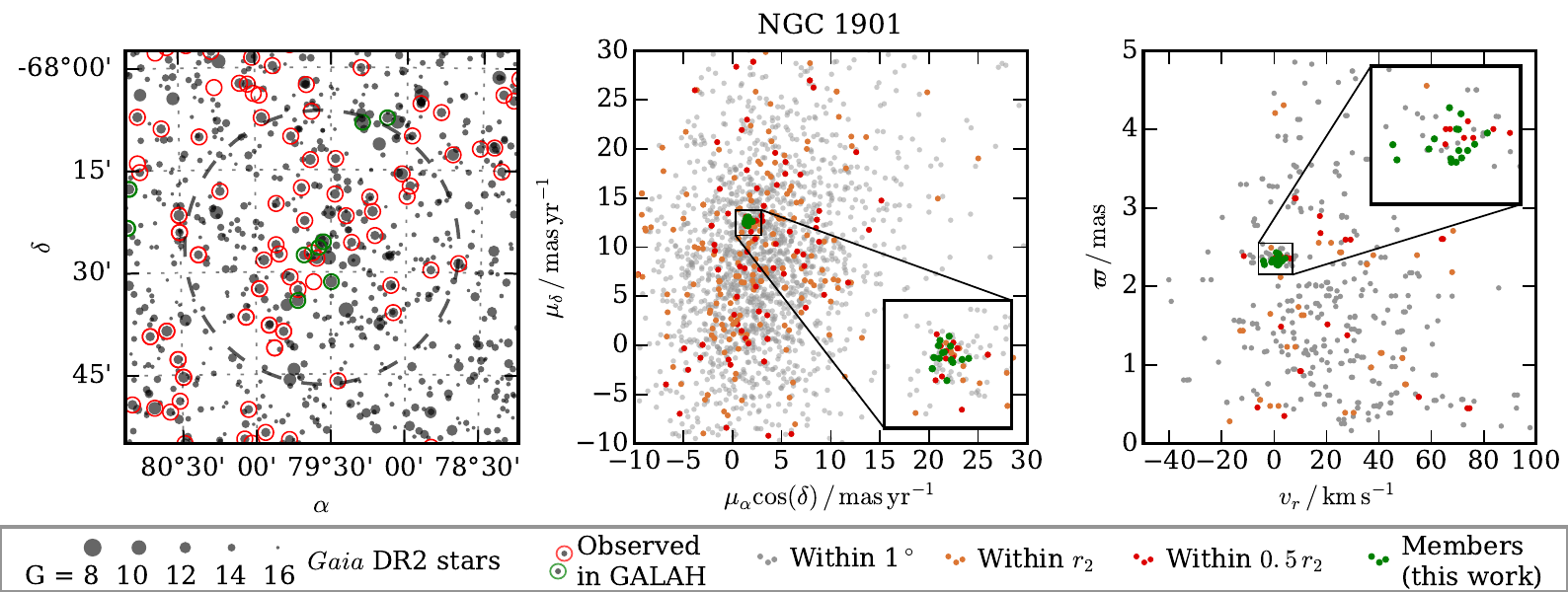}
\caption{Six dimensional space for stars in and around NGC~1901. Left: Position of \textit{Gaia} DR2 stars with magnitude G<16 (gray) with marked members observed in the GALAH survey (circled in green) and other stars observed in GALAH (circled in red). The dashed circle marks the cluster radius $r_2$. Middle: \textit{Gaia} DR2 proper motions for all stars inside a $1^\circ$ radius from the cluster centre (gray), stars inside $r_2$ (orange), stars inside $0.5\, r_2$ (red), and identified members (green). Right: Radial velocity and parallaxes for stars that have radial velocity measured either in \textit{Gaia} DR2 or in the GALAH survey. Same colours are used as in the middle panel. }
\label{fig:6d1}
\end{figure*}

Figures \ref{fig:6d1} and \ref{fig:6d2} show the six dimensional space ($\alpha$, $\delta$, $\mu_\alpha$, $\mu_\delta$, $\varpi$, $v_r$) for all five clusters. We choose to display observed parameters for various reasons; their uncertainties are mostly Gaussian and independent, which can not be said for derived parameters (actions, orbital parameters, ($U$,$V$,$W$) velocities, etc.), and when the radial velocity is not available we can still use the remaining five parameters. Also, a cluster is a hyperellipsoid in both observed and any derived space, so nothing can be gained from coordinate transformations. There is no common definition of a cluster (or a cluster remnant), but it makes sense that a necessary condition is that the cluster members form an overdensity in space, regardless of the dynamics of the cluster. This is another reason why measurables should be used, because they are already separated into three spatial dimensions and three velocities. In each figure we plot positions of stars on the sky ($\alpha$, $\delta$) in the left panel, while proper motions are plotted in the middle panel and the final two measurables ($\varpi$ and $v_r$) are plotted in the right panel. 

\begin{figure*}
\centering
\includegraphics[width=0.92\textwidth]{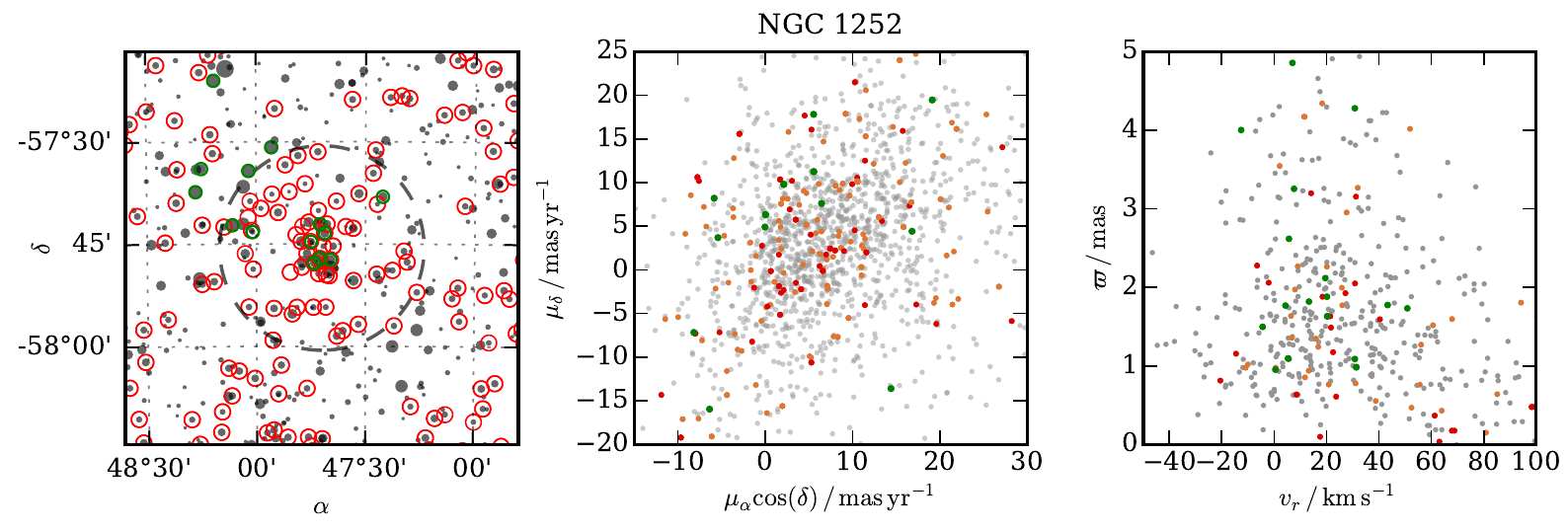}\\
\includegraphics[width=0.92\textwidth]{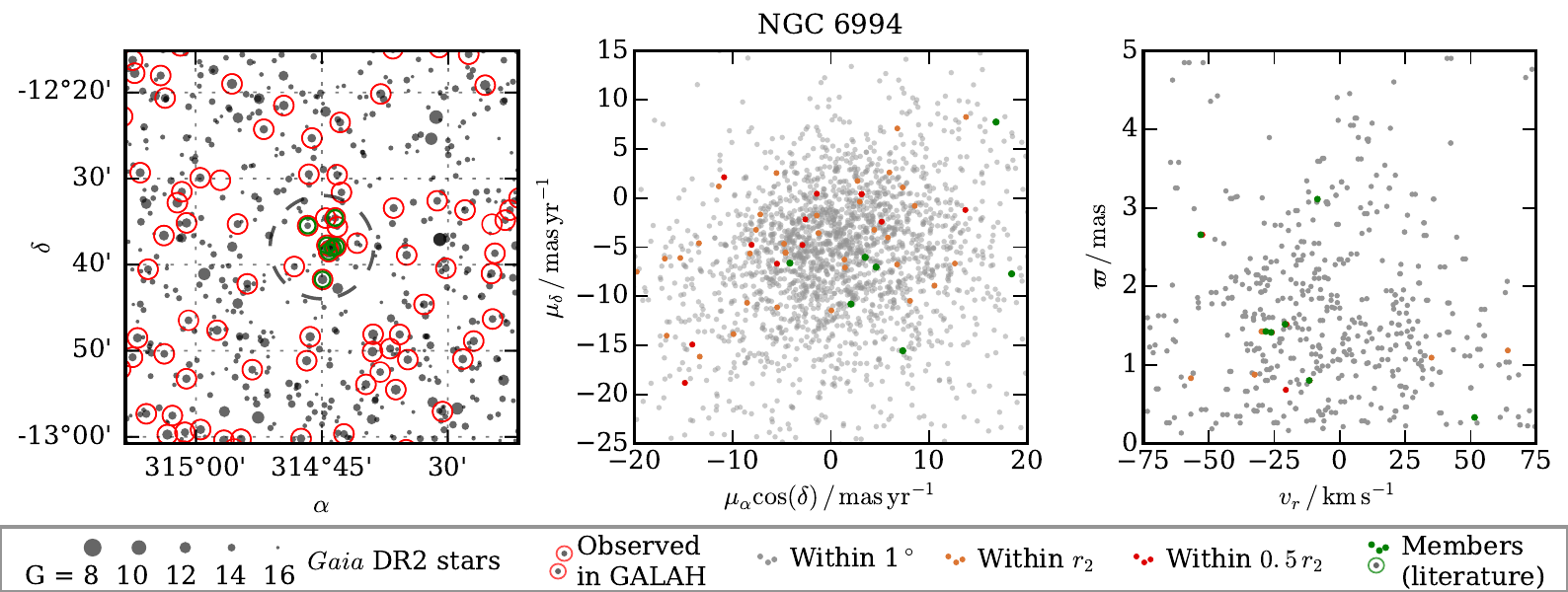}
\caption{Six dimensional space for stars in and around NGC~1252 (top row) and NGC~6994 (bottom row). Left: Position of \textit{Gaia} DR2 stars with magnitude G<16 (gray) with marked stars observed in the GALAH survey (circled in red) and faux members from the literature (circled in green). The dashed circle marks the cluster radius. Middle: \textit{Gaia} DR2 proper motions for all stars inside a $1^\circ$ radius from the cluster centre (gray), stars inside $r_2$ (orange), stars inside $0.5\, r_2$ (red), and faux members from the literature (green). Right: Radial velocity and parallaxes for stars that have radial velocity measured either in \textit{Gaia} DR2 or in the GALAH survey. Same colours are used as in the middle panel.}
\label{fig:6d2}
\end{figure*}

\begin{figure*}
\addtocounter{figure}{-1}
\centering
\includegraphics[width=0.92\textwidth]{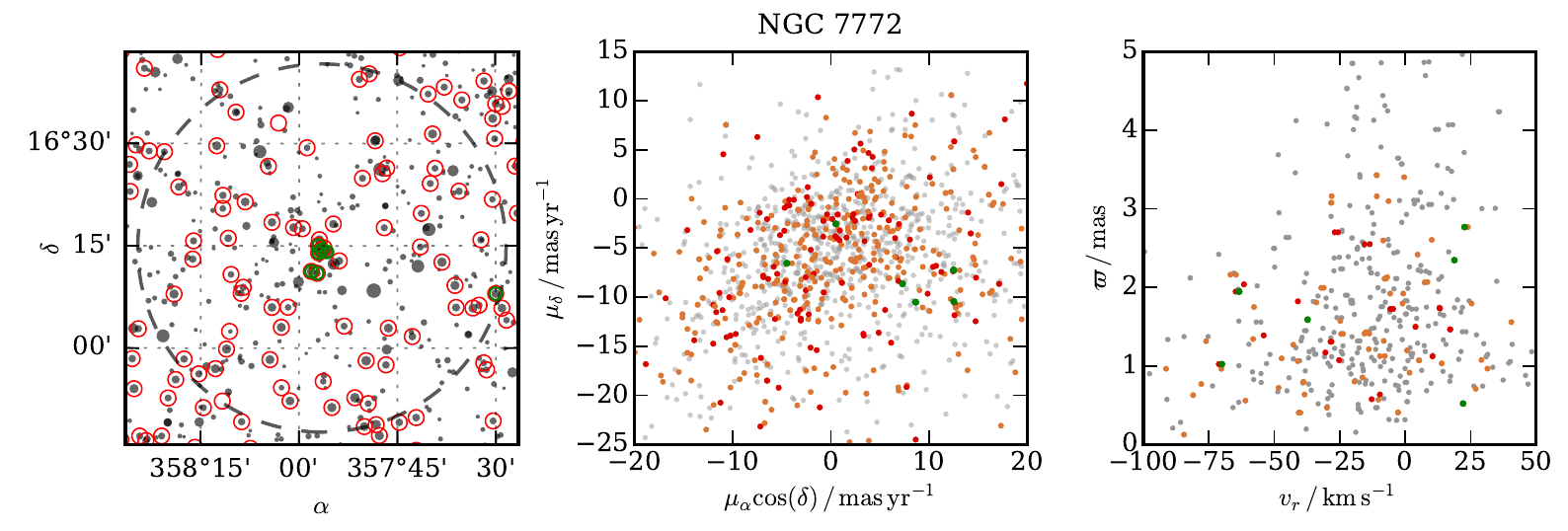}\\
\includegraphics[width=0.92\textwidth]{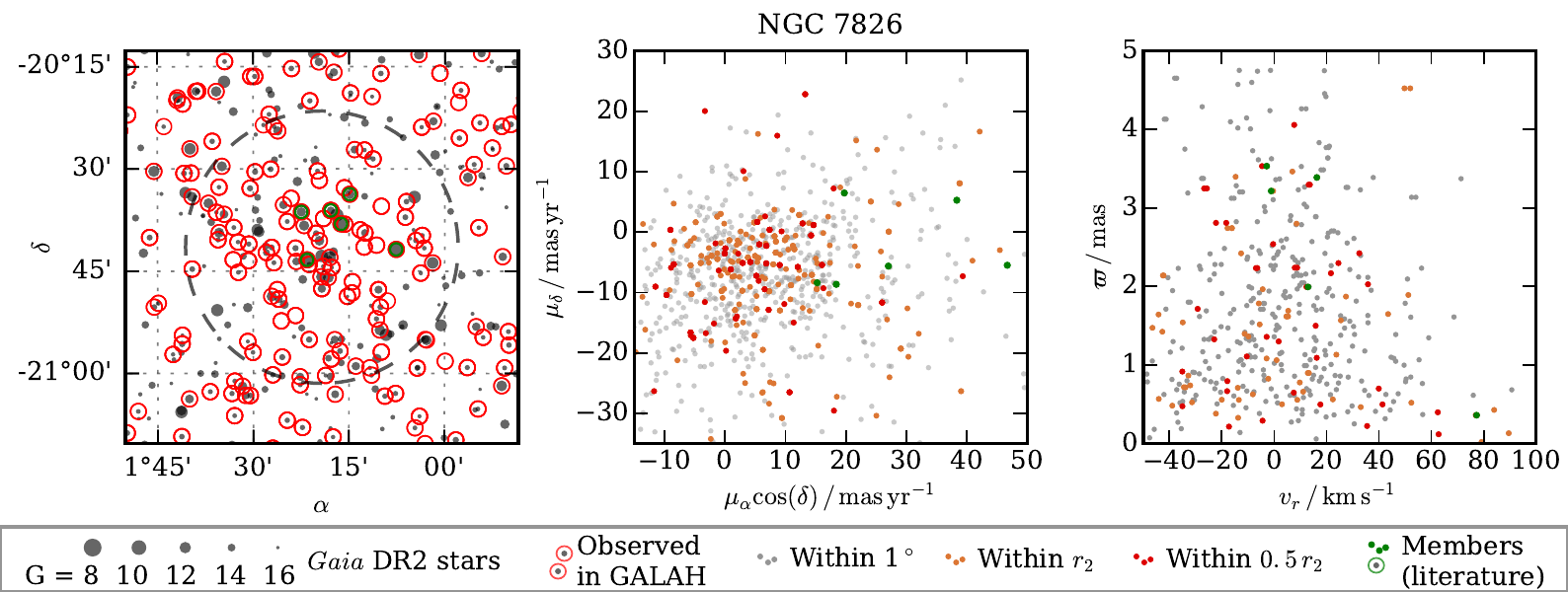}
\caption{Continued for NGC~7772 (top row) and NGC~7826 (bottom row).}
\end{figure*}

\begin{figure}
\centering
\includegraphics[width=\columnwidth]{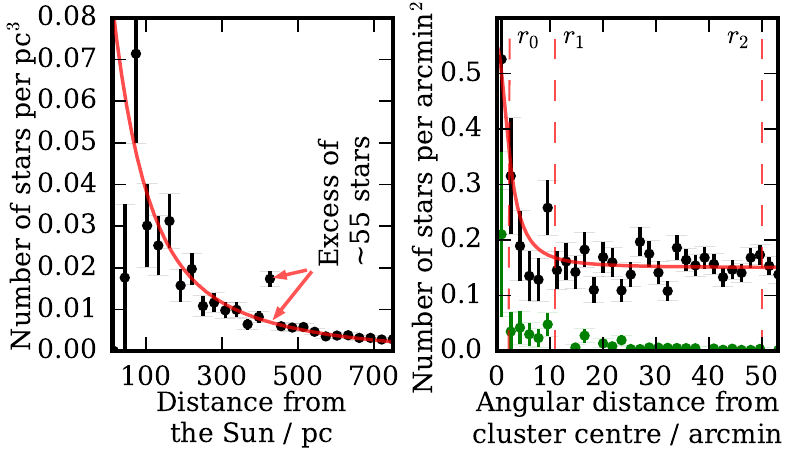}
\caption{Left: Density of stars as a function of distance from the Sun in the direction of NGC~1901. NGC~1901 produces a very significant ($5\sigma$) signal at 426~pc. Stars in a $1^\circ$ degree cone around the cluster centre are used for this plot. Right: Star density as a function of apparent distance from NGC~1901 centre. An overdensity above the background level (0.15 stars per $\mathrm{arcmin}^2$) is clearly detected. Black points show measurements for all stars with magnitude $\mathrm{G}<17$ and green points only show the density of the most probable members. Eyeballed radii $r_0$, $r_1$, and $r_2$ as defined in \citet{kharchenko12} are marked. Red curve is a fitted King model.}
\label{fig:ngc1901_3d}
\end{figure}

For NGC~1901, we clearly see clumping in both the proper-motion plane and the parallax-radial-velocity plane. An overdensity in all three spatial dimensions is illustrated in Figure~\ref{fig:ngc1901_3d}. The overdensity, however, is not obvious in the $(\alpha, \delta)$ plane. One would expect a similar clumping for the other four clusters, if they were real. Instead, we cannot find a single pair of stars (either among the literature members or all \textit{Gaia} stars), that are close together in all six dimensions, even though the overdensity in ($\alpha$, $\delta$) appears similar to that of NGC~1901.

\section{NGC~1901 analysis}
\label{sec:ngc1901members}

Since NGC~1901 is a real cluster, we provide here our own membership analysis. First, we performed a vague cut in the 6D space ($r<0.5^\circ$, $0.7<\mu_\alpha<2.5\, \mathrm{mas\,yr^{-1}}$, $11.7<\mu_\delta<13.5\, \mathrm{mas\,yr^{-1}}$, $1.8<\varpi<2.8\, \mathrm{mas}$, and $-2.0<v_r<7.0\, \mathrm{km\, s^{-1}}$) to isolate the most probable members. This yielded 80 stars, of which 20 have radial velocities. Most of these stars are members, therefore the mean position of these stars in the six-dimensional space is very close to the mean position of the cluster. We therefore used these stars to estimate the mean and spread in every dimension independently (there seems to be no correlation between different dimensions/parameters), which we used to perform a probability analysis. The probability for a star with given parameters ($\alpha$, $\delta$, $\mu_\alpha$, $\mu_\delta$, $\varpi$, and $v_r$) to be a member is described by a multivariate Gaussian centred on the mean values obtained by a vague cut. For stars with existing radial velocity measurement we produce two probabilities, one with radial velocity taken into account ($P_{6\mathrm{D}}$) and one without it ($P_{5\mathrm{D}}$). From the difference between these, we estimate that $\sim15$\% of highly probable members based only on the 5D analysis would have their probability reduced significantly if the radial velocity measurements were available. Membership probabilities are given in Table~\ref{tab:prob}. Re-fitting the multivariate Gaussian based on the improved membership probabilities changes the probabilities insignificantly and does not impact the rank of the most probable members.

From the most probable members we can calculate the mean parameters of NGC~1901, along with their uncertainties. In Table~\ref{tab:1901par} we provide the following parameters.  Positions $\alpha$, $\delta$ and $l$, $b$ give the cluster centre in celestial and galactic coordinates, respectively. Radii $r_0$, $r_1$, and $r_2$ are estimated visually as described in \citet{kharchenko12}. King's \citep{king62} cluster radius ($r_c$) and tidal radius ($r_t$) are fitted, although we were unable to measure the tidal radius with any meaningful confidence. Proper motions, $\mu_\alpha$ and $\mu_\delta$, are the mean values for the cluster and the uncertainties relate to the mean, not to the dispersion. The radial velocity ($v_r$) and the velocity dispersion ($\sigma_{v_r}$) are measured from a combination of \textit{Gaia} DR2 and GALAH radial velocities. The distance ($\mathrm{dist.}$) is calculated from the parallaxes ($\varpi$). A normal distribution for the parallax of each star was sampled and the distance calculated for every sample. The reported distance and its uncertainty are the mean and standard deviation of all the samples for all members. Distance uncertainty might be underestimated, if the parallaxes of individual stars have correlated errors. Iron and $\alpha$ abundances  $\left(\left[\mathrm{Fe} / \mathrm{H}\right],\  \left[\mathrm{\alpha} / \mathrm{Fe}\right]\right)$ are taken from GALAH and are based on 12 members only. Age ($\log\, t$) is calculated by isochrone fitting, assuming \textit{Gaia} reddening and extinction (see below).

\begin{table*}
\setlength{\tabcolsep}{0.8pt}
\caption{Final parameters for NGC~1901. Measurement uncertainties are given after the $\pm$ sign in the bottom row.}
\label{tab:1901par}
\centering
\scriptsize
\begin{tabular}{cccccccccccccccccc}
\hline\hline
$\alpha$ & $\delta$ & $l$ & $b$ & $r_0$ & $r_1$ & $r_2$ & $r_c$ & $r_t$ & $\mu_\alpha\, \cos(\delta)$ & $\mu_\delta$ & $v_r$ & $\varpi$ & dist. & $\sigma_{v_r}$ & $\left[\frac{\mathrm{Fe}}{\mathrm{H}}\right]$ & $\left[\frac{\mathrm{\alpha}}{\mathrm{Fe}}\right]$ & $\log\,t$\\[9pt]
$^\circ$ & $^\circ$ & $^\circ$ & $^\circ$ & $'$ & $'$ & $'$ & $'$ & $'$ & $\mathrm{mas\, y^{-1}}$ & $\mathrm{mas\, y^{-1}}$ & $\mathrm{km\, s^{-1}}$ & $\mathrm{mas}$ & pc & $\mathrm{km\, s^{-1}}$ & dex & dex & $\log\, yr$\\
\hline

$\begin{array}{r}79.490\\ \pm0.106\end{array}$ & $\begin{array}{r}-68.342\\ \pm0.031\end{array}$ & $\begin{array}{r}278.914\\ \pm0.039\end{array}$ & $\begin{array}{r}-33.644\\ \pm0.043\end{array}$ & $\begin{array}{r}2.2\\ \pm 0.5\end{array}$ & $\begin{array}{r}11.0\\ \pm 2.5\end{array}$ & $\begin{array}{r}50.0\\ \pm 12.0\end{array}$ & $\begin{array}{r}2.4\\ \pm 0.7 \end{array}$ & $>20$ & $\begin{array}{r}1.632\\ \pm0.030\end{array}$ & $\begin{array}{r}12.671\\ \pm0.018\end{array}$ & $\begin{array}{r}1.30\\ \pm0.28\end{array}$ & $\begin{array}{r}2.349\\ \pm0.009\end{array}$ & $\begin{array}{r}426.0\\ \pm2.0\end{array}$ & $\begin{array}{r}2.80\\ \pm0.27\end{array}$ & $\begin{array}{r}-0.32\\ \pm0.05\end{array}$ & $\begin{array}{r}0.11\\ \pm0.05\end{array}$ & $\begin{array}{r}8.26\\ \pm0.14\end{array}$\\
\hline
\end{tabular}
\normalsize
\end{table*}

\begin{figure}
\centering
\includegraphics[width=0.95\columnwidth]{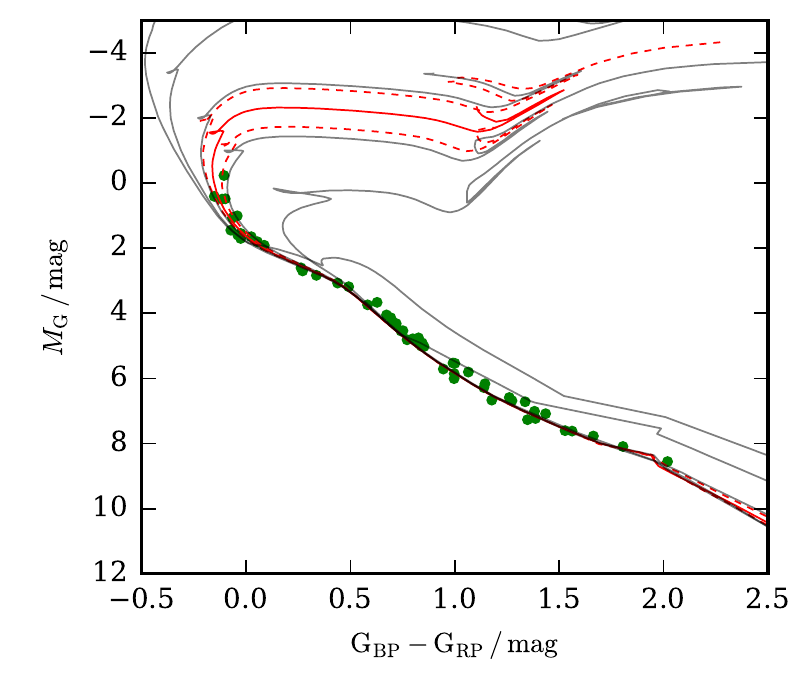}
\caption{NGC~1901 members on the color-magnitude diagram. Magnitudes of the stars are corrected for extinction and color excess. If either is unavailable in \textit{Gaia} DR2, we assumed a mean value for the cluster: $E\mathrm{(G_{BP}-G_{RP})}=0.20$ and $A_\mathrm{G}=0.32$. Padova isochrones \citep{marigo17} are plotted for $\log\, t=8.26$ (red solid curve, 182~Myr), and for $\log\, t=8.12$ and $\log\, t=8.40$ (red dashed curves), all with $\left[\mathrm{M} / \mathrm{H}\right]=-0.13$. Isochrones for $\log\, t=7.0$, $7.5$, $8.0$, $8.5,$ and $9.0$ are also plotted in gray. Varying the membership probability cut-off within reasonable values does not add or remove any turn-off stars. }
\label{fig:hr}
\end{figure}

Putting the most probable members onto the colour-magnitude diagram (Figure~\ref{fig:hr}) and fitting Padova isochrones \citep{marigo17} allows us to measure the age of the cluster. The precision of the calculated age is somewhat limited, as there are only a few stars close to the turn-off point. The reddening and colour excess as measured by \textit{Gaia} are higher than the literature values and the correlations between the age and reddening/extinction can increase the age uncertainty even further. From fitting the isochrones, we can also tell that there is a significant difference between the metallicity $\left(\left[\mathrm{M} / \mathrm{H}\right]=-0.13\right)$ of the best-matching isochrone and the iron abundance $\left(\left[\mathrm{Fe} / \mathrm{H}\right]=-0.32\right)$ we measured in the GALAH survey. Forcing the numbers to be the same by compensating for the metallicity by changing colour excess and extinction gives a significantly worse fit for the isochrones. This mismatch between GALAH and \textit{Gaia}~DR2 could be due to the difference between the two quantities. We also measured the abundance of other elements and they are close to or above solar values. A total metallicity (computed from the iron abundance, as well as abundances of other elements), which is not reported by the GALAH survey, should therefore be closer to the one measured from the isochrone fitting.

NGC~1901 has been referred to as an open cluster remnant in the literature \citep[e.g. by][]{carraro07}. While there is no clear consensus on the classification of an open cluster remnant, if an open cluster is sparsely populated with more than two-thirds of its initial stellar members lost, then a cluster like NGC~1901 is classed as a cluster remnant \citep{bica01}. However, as NGC~1901 is a rapidly dissolving cluster, it is not expected to survive another pass through the Galactic plane in around 18~Myr time (the last time it passed the Galactic plane was 26~Myr ago). We used \textit{galpy}\footnote{http://github.com/jobovy/galpy} \citep{bovy15} to calculate the orbit of individual stars in the cluster. Currently, the velocity dispersion of the cluster is 2.8~$\mathrm{km\, s^{-1}}$. By the time the cluster passes through the Galactic plane, the members will be spread out in a diameter $\sim$7 times larger than they are now. This is enough that the cluster will be undetectable with the approach used in this paper.

\cite{eggen96} referred to the NGC~1901 ``supercluster'', comprising a co-moving group of unbound stars associated with the cluster. The NGC~1901 supercluster has been proposed to be dynamically related to the Hyades supercluster (the unbound group of stars co-moving with the Hyades open cluster), due to the similar space motions of the two groups \citep{dehnen98}, and they are jointly referred to as Star Stream I by \cite{eggen96}. While the detection of the extended members of the NGC~1901 supercluster is beyond the scope of this paper, it is certainly plausible that the dispersed members of NGC~1901 are detectable within the Galactic motion space and could be used to gain insights into the dispersion mechanisms of the cluster.

\section{Discussion}
\label{sec:d}

Clusters cannot be simply split into high- and low-latitude, or sparse and rich clusters. There are a range of factors that could be specific to each cluster, depending on its initial mass, origin and evolutionary history. Furthermore, with typical cluster dispersion processes the transition is smooth and there are other clusters similar to the ones described in this paper among the 39 clusters in our data set, such as Blanco~1 and NGC~1817. We chose not to include such clusters here because they are slightly more populated and lie closer to the Galactic plane than the clusters discussed above. Also, the cluster sample in our data set was not observed with a clear selection function, apart from trying to cover as wide range of ages and metallicities as possible. As a result, it would not be reasonable to extrapolate from the results presented in this work to conclude that 4 out of 5 sparse high-latitude clusters are not real. However, a lesson learned is that the existence of sparse clusters should be double-checked, regardless of how reputable the respective cluster catalogues are. Surveys of high-latitude clusters \citep{bica01, schmeja14}, after the \textit{Gaia} parameters are included, will probably give a better picture.

\begin{figure*}
\centering
\includegraphics[width=0.95\textwidth]{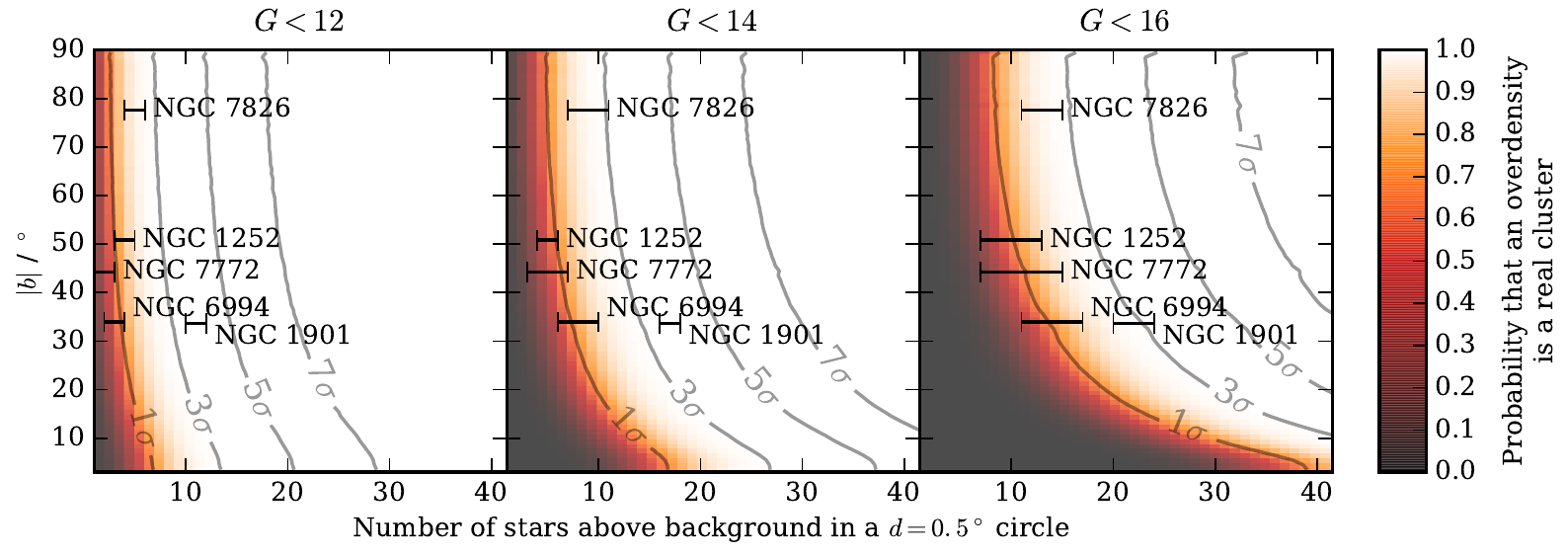}
\caption{Probability that an overdensity of $N$ stars above the background level in a $0.5^\circ$ diameter circle is a real cluster. Panels show the probability for different magnitude cuts between $G<12$ (left) and $G<16$ (right). The probability changes significantly with the background level. Here we only show the variation as a function of absolute galactic latitude $\left(|b|\right)$. Only minor differences in local density can be expected for regions around different clusters. Position of clusters studied in our work are indicated. For the nonexistent clusters we estimated the supposed number of stars from the literature sources. For all five clusters we assumed a radius of $0.25^\circ$, so all of them can be shown on the same plot.}
\label{fig:prob}
\end{figure*}

It is expected that some sparse clusters are not real. The rate of faux clusters can be estimated from mean star densities and the number of possible members in the aggregation. Figure~\ref{fig:prob} shows the probability as a function of the galactic latitude (a proxy for background star density) and the number of stars in the aggregation. We estimated the number of members of the four faux clusters from literature sources. The background star density is calculated from \textit{Gaia} DR2. All four faux clusters lie in the region where low probability clusters are expected to be found. 

For NGC~1901 we found more members than one would find using the same literature as for the four faux clusters, so the position of NGC~1901 on plots in Figure~\ref{fig:prob} with respect to the faux clusters might not be completely representative. NGC~1901 also lies in front of the LMC, so the background count in the $G<16$ panel is underestimated.

We can conclude that existence of the four false clusters has never been very plausible, since they were all discovered based on the star-counts only. Most probably, there are more long-known sparse clusters that will be disproved in the near future.

\textit{Gaia} parameters are obviously proving to be well suited for cluster membership analysis, as well as for finding new clusters \citep{koposov17, castro18, torrealba18, gaudin18, dias18}. The fraction of reported clusters that are not real will probably be significantly less than in the past, but even in the \textit{Gaia} era we can expect to find low-probability clusters that should be treated with caution. The same holds for membership probabilities. We show that positions, proper motions and distances are not enough and the whole 6D information must be used to find members with great certainty.

\section*{Acknowledgements}

JK is supported by a Discovery Project grant from the Australian Research Council (DP150104667) awarded to J. Bland-Hawthorn and T. Bedding. SB acknowledges funds from the Alexander von Humboldt Foundation in the framework of the Sofja Kovalevskaja Award endowed by the Federal Ministry of Education and Research. SLM acknowledges support from the Australian Research Council through grant DP180101791. Parts of this research were conducted by the Australian Research Council Centre of Excellence for All Sky Astrophysics in Three Dimensions (ASTRO 3D), through project number CE170100013. GT, K\v{C}, and TZ acknowledge the financial support from the Slovenian Research Agency (research core funding No. P1-0188).

%%%%%%%%%%%%%%%%%%%%%%%%%%%%%%%%%%%%%%%%%%%%%%%%%%

%%%%%%%%%%%%%%%%%%%% REFERENCES %%%%%%%%%%%%%%%%%%

% The best way to enter references is to use BibTeX:

\bibliographystyle{mnras}
\bibliography{bib}

%%%%%%%%%%%%%%%%%%%%%%%%%%%%%%%%%%%%%%%%%%%%%%%%%%

%%%%%%%%%%%%%%%%% APPENDICES %%%%%%%%%%%%%%%%%%%%%

\clearpage
\appendix

\section{Literature members for NGC~1252, NGC~6994, NGC~7772, and NGC~7826}

Tables \ref{tab:members1252} to \ref{tab:members7826} list members given in the literature for four disproved clusters. Only stars we were able to cross-match with \textit{Gaia}~DR2 are listed. These stars are clearly not real members, but we use them to rest our case in Figure~\ref{fig:6d2}.

\begin{table}
\setlength{\tabcolsep}{4.2pt}
\caption{Coordinates, proper motions, parallax and radial velocity (where known) of NGC~1252 members from the literature. Not a single pair of stars can be found with matching parameters. Between all four sources the magnitudes of the stars extend from $V=6.62$ to $V=17.97$.}
\scriptsize
\label{tab:members1252}
\begin{tabular}{cccccc}
\hline\hline
$\alpha$ & $\delta$ & $\mu_\alpha\cos(\delta)$ & $\mu_\delta$ & $\varpi$ & $v_r$\\
$^\circ$ & $^\circ$ & $\mathrm{mas\, y^{-1}}$ & $\mathrm{mas\, y^{-1}}$ & $\mathrm{mas}$ & $\mathrm{km\, s^{-1}}$ \\
\hline\\[-0.15cm]
\multicolumn{6}{c}{Members$^\ddagger$ in \citet{kharchenko13}}\\[0.1cm]
47.4215 & -57.6347 & -0.03$\pm$0.05 & 4.90$\pm$0.04 & 0.95$\pm$0.03 & 0.5$\pm$0.6$^*$ \\
47.6764 & -57.7014 & -2.96$\pm$0.03 & 15.57$\pm$0.03 & 3.20$\pm$0.02 & \\
47.2933 & -57.7678 & -1.07$\pm$0.06 & -13.19$\pm$0.05 & 1.08$\pm$0.03 & \\
48.0644 & -57.6689 & -4.70$\pm$0.05 & 3.90$\pm$0.05 & 1.39$\pm$0.03 & \\
47.7155 & -57.6683 & 1.57$\pm$0.05 & -1.88$\pm$0.04 & 0.61$\pm$0.03 & \\
47.9065 & -57.5849 & 9.76$\pm$0.15 & 3.23$\pm$0.14 & 1.26$\pm$0.07 & \\
48.0170 & -57.7205 & 2.12$\pm$0.05 & 9.79$\pm$0.04 & 0.98$\pm$0.02 & 31.4$\pm$0.3$^\dagger$ \\
47.3650 & -57.6324 & -3.59$\pm$0.10 & 12.80$\pm$0.09 & 1.72$\pm$0.05 & \\
47.7089 & -57.7856 & 0.04$\pm$0.05 & 6.32$\pm$0.04 & 1.63$\pm$0.03 & 20.3$\pm$0.8$^*$ \\
47.3982 & -57.8135 & 6.42$\pm$0.04 & 4.39$\pm$0.03 & 0.76$\pm$0.02 & \\
47.9903 & -57.6525 & 4.39$\pm$0.14 & 9.04$\pm$0.14 & 1.49$\pm$0.07 & \\[0.15cm]
\multicolumn{6}{c}{Members in \citet{delafuente13}}\\[0.1cm]
47.6605 & -57.7887 & 6.44$\pm$0.05 & 7.61$\pm$0.04 & 1.50$\pm$0.02 & -4.4$\pm$1.1$^*$ \\
47.7356 & -57.7966 & 5.55$\pm$0.04 & 11.25$\pm$0.03 & 1.09$\pm$0.02 & 5.3$\pm$0.2$^\dagger$ \\
47.7498 & -57.6912 & 9.95$\pm$0.08 & 9.81$\pm$0.07 & 0.85$\pm$0.05 & \\
47.4953 & -57.7548 & 3.06$\pm$0.10 & 10.20$\pm$0.09 & 0.66$\pm$0.05 & \\
47.6336 & -57.6454 & 3.49$\pm$0.05 & 5.74$\pm$0.05 & 0.81$\pm$0.03 & \\
47.9003 & -57.6730 & 13.01$\pm$0.03 & 6.05$\pm$0.03 & 2.26$\pm$0.02 & \\
48.0170 & -57.7205 & 2.12$\pm$0.05 & 9.79$\pm$0.04 & 0.98$\pm$0.02 & 31.4$\pm$0.3$^\dagger$ \\
48.0104 & -57.6817 & 5.09$\pm$0.07 & -0.90$\pm$0.07 & 0.25$\pm$0.04 & \\[0.15cm]
\multicolumn{6}{c}{Members in \citet{pavani01}}\\[0.1cm]
47.7089 & -57.7856 & 0.04$\pm$0.05 & 6.32$\pm$0.04 & 1.63$\pm$0.03 & 20.3$\pm$0.8$^*$ \\
47.6868 & -57.7234 & -21.39$\pm$0.04 & -44.09$\pm$0.04 & 1.88$\pm$0.02 & 20.1$\pm$1.5$^*$ \\
47.7680 & -57.7731 & -17.57$\pm$0.05 & -18.16$\pm$0.05 & 2.05$\pm$0.03 & \\
47.7356 & -57.7966 & 5.55$\pm$0.04 & 11.25$\pm$0.03 & 1.09$\pm$0.02 & 5.3$\pm$0.2$^\dagger$ \\
47.7498 & -57.7446 & 31.21$\pm$0.03 & 17.01$\pm$0.03 & 1.77$\pm$0.02 & 43.3$\pm$4.4$^*$ \\
47.6764 & -57.7014 & -2.96$\pm$0.03 & 15.57$\pm$0.03 & 3.20$\pm$0.02 & \\
47.7083 & -57.7021 & 4.05$\pm$0.07 & -69.89$\pm$0.06 & 11.87$\pm$0.04 & 54.0$\pm$0.3$^*$ \\[0.15cm]
\multicolumn{6}{c}{Members in \citet{bouchet83}}\\[0.1cm]
47.7083 & -57.7021 & 4.05$\pm$0.07 & -69.89$\pm$0.06 & 11.87$\pm$0.04 & 54.0$\pm$0.3$^*$ \\
48.1091 & -57.7038 & 35.61$\pm$0.04 & 1.30$\pm$0.04 & 1.73$\pm$0.02 & 50.8$\pm$0.2$^*$ \\
48.2607 & -57.8339 & 1.19$\pm$0.07 & -4.57$\pm$0.08 & 1.08$\pm$0.05 & \\
48.2768 & -57.6225 & 5.54$\pm$0.05 & 17.82$\pm$0.06 & 4.28$\pm$0.03 & 30.8$\pm$0.5$^*$ \\
48.2507 & -57.5659 & 16.81$\pm$0.04 & 4.41$\pm$0.04 & 1.09$\pm$0.02 & 30.8$\pm$0.5$^*$ \\
48.0351 & -57.5712 & 43.18$\pm$0.06 & 15.34$\pm$0.06 & 3.25$\pm$0.03 & 7.6$\pm$0.3$^*$ \\
47.9307 & -57.5136 & -5.85$\pm$0.04 & 8.20$\pm$0.04 & 2.12$\pm$0.02 & 19.5$\pm$0.2$^*$ \\
48.1921 & -57.3504 & -5.42$\pm$0.06 & 3.69$\pm$0.06 & 4.00$\pm$0.03 & -12.6$\pm$0.5$^*$ \\
47.7845 & -57.1932 & -3.39$\pm$0.09 & -26.14$\pm$0.10 & 4.86$\pm$0.05 & 7.0$\pm$0.5$^*$ \\
47.8202 & -57.1590 & 19.11$\pm$0.24 & 19.46$\pm$0.23 & 1.77$\pm$0.13 & 4.3$\pm$0.3$^*$ \\
47.9257 & -56.9778 & -8.18$\pm$0.04 & -7.17$\pm$0.04 & 1.82$\pm$0.02 & 13.3$\pm$0.2$^*$ \\
47.0998 & -57.0100 & 14.40$\pm$0.04 & -13.60$\pm$0.05 & 2.62$\pm$0.03 & 5.6$\pm$0.2$^*$ \\
48.7182 & -57.8377 & -6.38$\pm$0.08 & -15.96$\pm$0.08 & 2.82$\pm$0.05 & 105.3$\pm$0.3$^*$ \\
\hline\\[-0.3cm]
\end{tabular}
\\$^\ddagger$ only members with probability>0.8\\[-0.1cm]
$^*$ $v_r$ from \textit{Gaia} DR2\\[-0.1cm]
$^\dagger$ $v_r$ from GALAH
\normalsize
\end{table}

\begin{table}
\setlength{\tabcolsep}{4.2pt}
\caption{Coordinates, proper motions, parallax and radial velocity (where known) of NGC~6994 members from the literature. Not a single pair of stars can be found with matching parameters. The magnitudes of the stars extend from $V=10.35$ to $V=19.53$.}
\scriptsize
\label{tab:members6994}
\begin{tabular}{cccccc}
\hline\hline
$\alpha$ & $\delta$ & $\mu_\alpha\cos(\delta)$ & $\mu_\delta$ & $\varpi$ & $v_r$\\
$^\circ$ & $^\circ$ & $\mathrm{mas\, y^{-1}}$ & $\mathrm{mas\, y^{-1}}$ & $\mathrm{mas}$ & $\mathrm{km\, s^{-1}}$ \\
\hline\\[-0.15cm]
\multicolumn{6}{c}{Members in \citet{bassino00}}\\[0.1cm]
314.7366 & -12.6418 & 7.33$\pm$0.07 & -15.55$\pm$0.05 & 1.41$\pm$0.04 & -26.2$\pm$0.2$^*$ \\
314.7401 & -12.6294 & 18.41$\pm$0.07 & -7.72$\pm$0.06 & 2.66$\pm$0.05 & -53.1$\pm$0.7$^*$ \\
314.7283 & -12.6345 & 2.05$\pm$0.09 & -10.80$\pm$0.07 & 3.11$\pm$0.06 & -8.5$\pm$0.6$^*$ \\
314.7222 & -12.6318 & 4.61$\pm$0.07 & -7.03$\pm$0.05 & 1.52$\pm$0.04 & -20.8$\pm$0.6$^*$ \\
314.8046 & -12.6706 & -21.38$\pm$0.05 & -9.39$\pm$0.03 & 1.18$\pm$0.03 & \\
314.7781 & -12.5914 & 3.50$\pm$0.05 & -6.02$\pm$0.03 & 0.80$\pm$0.03 & -11.6$\pm$0.1$^\dagger$ \\
314.7776 & -12.5777 & -11.40$\pm$0.07 & 1.19$\pm$0.05 & 1.91$\pm$0.05 & \\
314.7423 & -12.5768 & -4.60$\pm$0.04 & -5.57$\pm$0.03 & 0.83$\pm$0.03 & \\
314.7237 & -12.5765 & -4.16$\pm$0.04 & -6.62$\pm$0.03 & 0.33$\pm$0.03 & 51.5$\pm$0.1$^\dagger$ \\
314.7378 & -12.5785 & -5.53$\pm$0.35 & 2.55$\pm$0.23 & 1.78$\pm$0.23 & \\
314.7487 & -12.6955 & 16.82$\pm$0.07 & 7.75$\pm$0.04 & 1.43$\pm$0.04 & -28.2$\pm$1.6$^*$ \\
\hline\\[-0.3cm]
\end{tabular}
\\$^*$ $v_r$ from \textit{Gaia} DR2\\[-0.1cm]
$^\dagger$ $v_r$ from GALAH
\normalsize
\end{table}

\begin{table}
\setlength{\tabcolsep}{4.2pt}
\caption{Coordinates, proper motions, parallax and radial velocity (where known) of NGC~7772 members from the literature. Not a single pair of stars can be found with matching parameters. Between both sources the magnitudes of the stars extend from $V=11.08$ to $V=18.00$.}
\scriptsize
\label{tab:members7772}
\begin{tabular}{cccccc}
\hline\hline
$\alpha$ & $\delta$ & $\mu_\alpha\cos(\delta)$ & $\mu_\delta$ & $\varpi$ & $v_r$\\
$^\circ$ & $^\circ$ & $\mathrm{mas\, y^{-1}}$ & $\mathrm{mas\, y^{-1}}$ & $\mathrm{mas}$ & $\mathrm{km\, s^{-1}}$ \\
\hline\\[-0.15cm]
\multicolumn{6}{c}{Members$^\ddagger$ in \citet{kharchenko13}}\\[0.1cm]
357.9917 & 16.2920 & 17.13$\pm$0.05 & -8.65$\pm$0.02 & 1.39$\pm$0.03 & \\
357.8054 & 16.1146 & 6.94$\pm$0.34 & -2.56$\pm$0.14 & 1.30$\pm$0.16 & \\
357.5001 & 16.1331 & 12.52$\pm$0.09 & -10.44$\pm$0.04 & 2.77$\pm$0.04 & 22.7$\pm$1.7$^*$ \\
358.2582 & 16.4424 & 18.28$\pm$0.13 & -14.25$\pm$0.06 & 1.03$\pm$0.07 & \\
357.9434 & 16.2490 & 12.45$\pm$0.05 & -7.38$\pm$0.02 & 1.88$\pm$0.03 & \\
357.9288 & 16.2352 & 7.30$\pm$0.08 & -8.65$\pm$0.04 & 0.52$\pm$0.04 & 22.1$\pm$0.3$^*$ \\
357.9052 & 15.9786 & 11.59$\pm$0.10 & -7.66$\pm$0.04 & 1.05$\pm$0.05 & \\
358.0828 & 16.2812 & 10.42$\pm$0.30 & -6.91$\pm$0.18 & 1.53$\pm$0.15 & \\
357.9428 & 16.2398 & 12.50$\pm$0.07 & -7.25$\pm$0.04 & 1.95$\pm$0.04 & -63.4$\pm$0.1$^\dagger$ \\
358.3233 & 16.0572 & 16.19$\pm$0.09 & -9.59$\pm$0.04 & 1.03$\pm$0.05 & \\[0.15cm]
\multicolumn{6}{c}{Members in \citet{carraro02}}\\[0.1cm]
357.9428 & 16.2398 & 12.50$\pm$0.07 & -7.25$\pm$0.04 & 1.95$\pm$0.04 & -63.4$\pm$0.1$^\dagger$ \\
357.9690 & 16.1878 & 6.75$\pm$0.16 & -38.26$\pm$0.08 & 5.48$\pm$0.09 & -20.1$\pm$1.2$^*$ \\
357.9129 & 16.3036 & -16.76$\pm$0.06 & -31.33$\pm$0.03 & 2.04$\pm$0.03 & \\
357.9506 & 16.2513 & 8.65$\pm$0.06 & -10.52$\pm$0.03 & 1.02$\pm$0.03 & -69.9$\pm$1.3$^*$ \\
357.9434 & 16.2490 & 12.45$\pm$0.05 & -7.38$\pm$0.02 & 1.88$\pm$0.03 & \\
357.8977 & 16.2136 & -6.48$\pm$0.05 & -18.50$\pm$0.02 & 0.64$\pm$0.03 & \\
357.9503 & 16.2333 & 0.52$\pm$0.05 & -2.54$\pm$0.03 & 2.34$\pm$0.03 & 18.8$\pm$0.2$^\dagger$ \\
357.9917 & 16.2920 & 17.13$\pm$0.05 & -8.65$\pm$0.02 & 1.39$\pm$0.03 & \\
357.9367 & 16.2454 & 12.56$\pm$0.05 & 5.86$\pm$0.03 & 1.50$\pm$0.03 & \\
357.9538 & 16.1833 & -4.48$\pm$0.05 & -6.55$\pm$0.03 & 1.59$\pm$0.03 & -37.2$\pm$0.1$^\dagger$ \\
357.9056 & 16.3006 & 2.74$\pm$0.08 & 0.49$\pm$0.04 & 0.48$\pm$0.04 & \\
\hline\\[-0.3cm]
\end{tabular}
\\$^\ddagger$ only members with probability>0.8\\[-0.1cm]
$^*$ $v_r$ from \textit{Gaia} DR2\\[-0.1cm]
$^\dagger$ $v_r$ from GALAH
\normalsize
\end{table}

\begin{table}
\setlength{\tabcolsep}{4.2pt}
\caption{Coordinates, proper motions, parallax and radial velocity (where known) of NGC~7826 members from the literature. Not a single pair of stars can be found with matching parameters. The magnitudes of the stars extend from $V=9.72$ to $V=16.67$.}
\scriptsize
\label{tab:members7826}
\begin{tabular}{cccccc}
\hline\hline
$\alpha$ & $\delta$ & $\mu_\alpha\cos(\delta)$ & $\mu_\delta$ & $\varpi$ & $v_r$\\
$^\circ$ & $^\circ$ & $\mathrm{mas\, y^{-1}}$ & $\mathrm{mas\, y^{-1}}$ & $\mathrm{mas}$ & $\mathrm{km\, s^{-1}}$ \\
\hline\\[-0.15cm]
\multicolumn{6}{c}{Members$^\ddagger$ in \citet{dias02, dias14}}\\[0.1cm]
1.3897 & -20.8588 & 12.26$\pm$0.12 & 2.83$\pm$0.07 & 0.72$\pm$0.07 & \\
1.1267 & -20.6970 & 46.70$\pm$0.09 & -5.49$\pm$0.06 & 3.22$\pm$0.06 & -1.1$\pm$0.2$^*$ \\
1.2085 & -20.6567 & 25.99$\pm$0.06 & -11.67$\pm$0.04 & 1.30$\pm$0.03 & \\
1.2711 & -20.6347 & 15.25$\pm$0.07 & -8.40$\pm$0.05 & 0.36$\pm$0.04 & 77.3$\pm$0.1$^\dagger$ \\
1.2949 & -20.7413 & 14.65$\pm$0.10 & 1.17$\pm$0.06 & 1.18$\pm$0.05 & \\
1.2976 & -20.6020 & 19.74$\pm$0.05 & 6.47$\pm$0.03 & 0.79$\pm$0.03 & -17.8$\pm$0.1$^\dagger$ \\
1.3595 & -20.7226 & 38.34$\pm$0.12 & 5.27$\pm$0.06 & 3.39$\pm$0.05 & 16.3$\pm$0.7$^*$ \\
1.3745 & -20.6056 & 27.03$\pm$0.07 & -5.63$\pm$0.04 & 3.53$\pm$0.04 & -2.9$\pm$0.1$^\dagger$ \\
1.4658 & -20.7048 & 16.05$\pm$0.05 & -5.44$\pm$0.04 & 0.92$\pm$0.04 & \\
1.2480 & -20.5623 & 18.41$\pm$0.07 & -8.65$\pm$0.05 & 1.99$\pm$0.04 & 12.9$\pm$1.1$^*$ \\
1.4210 & -20.5883 & 12.84$\pm$0.08 & 1.43$\pm$0.04 & 2.54$\pm$0.04 & \\
\hline\\[-0.3cm]
\end{tabular}
\\$^\ddagger$ only members with probability>0.9\\[-0.1cm]
$^*$ $v_r$ from \textit{Gaia} DR2\\[-0.1cm]
$^\dagger$ $v_r$ from GALAH
\normalsize
\end{table}

\section{List of probable NGC~1901 members}

\begin{figure}
\centering
\includegraphics[width=\columnwidth]{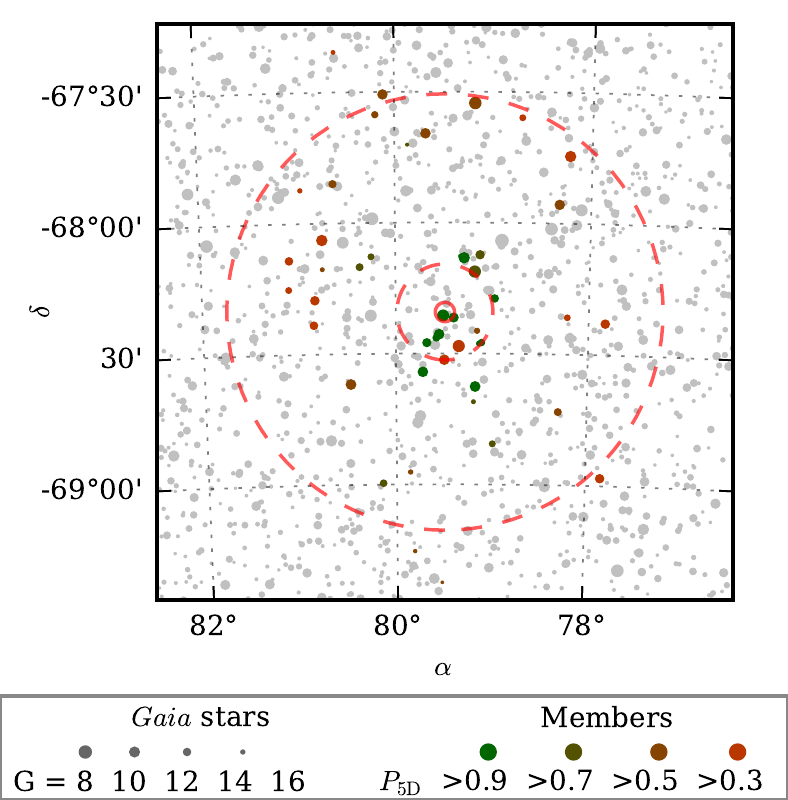}
\caption{Positions of the most probable NGC~1901 members. All \textit{Gaia}~DR2 stars with $G<16$ are plotted in grey and the members are plotted in colour. Radii $r_0$, $r_1$, and $r_2$ are marked with red circles.}
\label{fig:a_pos}
\end{figure}

In Table~\ref{tab:prob} we list the most probable NGC~1901 members. Note that only stars in the \textit{Gaia}~DR2 are included. Figure~\ref{fig:a_pos} shows the position of the most probable members on the sky.

\begin{table*}
\caption{List of most probable NGC~1901 members ordered by probability. Probability $P_{5\mathrm{D}}$ is given to all stars in \textit{Gaia} DR2 with known position, proper motions and parallax. The uncertainty of all measurements is properly taken into account. Stars with known radial velocity are given another probability ($P_{6\mathrm{D}}$) based on their radial velocity as well. Stars with GALAH radial velocities are marked with an asterisk.}
\label{tab:prob}
\scriptsize
\setlength{\tabcolsep}{4pt}
\begin{tabular}{ccccccccccc}
\hline\hline
\textit{Gaia} DR2 source id & $\alpha$ & $\delta$ & $r$ & $\mu_\alpha\, \cos(\delta)$ & $\mu_\delta$ & $\varpi$ & $G$ & $v_r$ & $P_{5\mathrm{D}}$ & $P_{6\mathrm{D}}$\\\\[-1pt]
 & $^\circ$ & $^\circ$ & $^\circ$ & $\mathrm{mas\, y^{-1}}$ & $\mathrm{mas\, y^{-1}}$ & $\mathrm{mas}$ & $\mathrm{mag}$ & $\mathrm{km\, s^{-1}}$ & & \\
\hline

4658338537313572096 & 79.5799 & -68.4403 & 0.1037 & 1.490 & 12.736 & 2.368 & 12.48 & 1.43$^*$ & 0.980 & 0.979 \\
4658335518014051712 & 79.6781 & -68.4590 & 0.1360 & 1.666 & 12.592 & 2.324 & 11.49 & 3.01$^*$ & 0.972 & 0.938 \\
4658765564465430272 & 79.3346 & -68.1278 & 0.2218 & 1.727 & 12.684 & 2.333 & 14.78 &   & 0.964 & \\
4658338747829382528 & 79.5499 & -68.4267 & 0.0875 & 1.425 & 12.620 & 2.340 & 10.33 & -0.49$^*$ & 0.963 & 0.926 \\
4658340053499327360 & 79.3992 & -68.3630 & 0.0395 & 1.792 & 12.825 & 2.388 & 10.81 & -11.58$^*$ & 0.960 & 0.124 \\
4658714677689040640 & 79.5087 & -68.3554 & 0.0150 & 1.709 & 12.942 & 2.331 & 9.21 &   & 0.943 & \\
4658384751164454016 & 79.1349 & -68.4621 & 0.1775 & 1.558 & 12.617 & 2.360 & 13.52 &   & 0.943 & \\
4658384755525236864 & 79.1097 & -68.4584 & 0.1821 & 1.703 & 12.647 & 2.357 & 12.72 & 6.07 & 0.931 & 0.704 \\
4658333422070733312 & 79.7205 & -68.5703 & 0.2435 & 1.652 & 12.869 & 2.323 & 10.37 & -4.09$^*$ & 0.927 & 0.647 \\
4658726218241467008 & 79.6766 & -68.0353 & 0.3144 & 1.819 & 12.558 & 2.361 & 10.40 &   & 0.912 & \\
4658765736264109312 & 79.2926 & -68.1346 & 0.2199 & 1.700 & 12.537 & 2.282 & 9.74 & 0.95$^*$ & 0.910 & 0.909 \\
4658378570770738944 & 79.2189 & -68.5038 & 0.1900 & 1.404 & 12.759 & 2.367 & 16.14 &   & 0.899 & \\
4658329298899347328 & 79.1740 & -68.6263 & 0.3070 & 1.655 & 12.778 & 2.290 & 10.21 &   & 0.891 & \\
4658388260218178048 & 78.9761 & -68.2895 & 0.1970 & 1.671 & 12.729 & 2.398 & 11.98 & 54.53$^*$ & 0.887 & 0.000 \\
4658766114221175040 & 79.1298 & -68.1230 & 0.2565 & 1.888 & 12.597 & 2.367 & 11.34 & 1.61$^*$ & 0.878 & 0.877 \\
4658328680424119552 & 79.1905 & -68.6844 & 0.3596 & 1.834 & 12.825 & 2.391 & 14.23 &   & 0.874 & \\
4658751030294843776 & 79.5615 & -67.9204 & 0.4225 & 1.749 & 12.950 & 2.382 & 15.04 &   & 0.786 & \\
4658709970404987904 & 79.9753 & -68.4676 & 0.2184 & 1.453 & 12.870 & 2.434 & 15.89 &   & 0.762 & \\
4658349364997640064 & 78.9891 & -68.8448 & 0.5350 & 1.885 & 12.586 & 2.309 & 13.02 &   & 0.757 & \\
4658764499313514112 & 79.1826 & -68.1872 & 0.1921 & 1.911 & 12.461 & 2.443 & 9.03 &   & 0.737 & \\
4658760822820870144 & 79.8711 & -67.7024 & 0.6553 & 1.586 & 12.884 & 2.409 & 14.73 &   & 0.727 & \\
4658748109717269120 & 80.1748 & -67.9189 & 0.4940 & 1.447 & 12.760 & 2.395 & 16.20 &   & 0.723 & \\
4658733335028239232 & 80.2483 & -68.1295 & 0.3524 & 1.458 & 12.684 & 2.430 & 13.04 &   & 0.716 & \\
4658383033177728512 & 78.8183 & -68.3973 & 0.2537 & 1.783 & 12.413 & 2.330 & 16.77 &   & 0.696 & \\
4658732265555421952 & 80.3665 & -68.1686 & 0.3681 & 1.416 & 12.418 & 2.375 & 12.39 & 13.68$^*$ & 0.678 & 0.102 \\
4658286069991771392 & 80.1431 & -68.9948 & 0.6947 & 1.648 & 12.546 & 2.349 & 12.56 & -1.51 & 0.678 & 0.615 \\
4658764877270576000 & 79.0623 & -68.1334 & 0.2620 & 1.703 & 13.032 & 2.426 & 8.79 &   & 0.672 & \\
4658332631796592768 & 79.7948 & -68.5810 & 0.2639 & 1.478 & 12.252 & 2.385 & 9.83 &   & 0.659 & \\
4658761887972710144 & 79.6859 & -67.6594 & 0.6865 & 1.438 & 12.987 & 2.285 & 10.41 & 1.87$^*$ & 0.650 & 0.648 \\
4658385305280833152 & 79.1563 & -68.4135 & 0.1422 & 1.372 & 13.111 & 2.354 & 13.53 &   & 0.620 & \\
4658765736264108800 & 79.2927 & -68.1340 & 0.2205 & 1.372 & 12.756 & 2.234 & 12.85 &   & 0.617 & \\
4658758963072125312 & 79.8855 & -67.7579 & 0.6025 & 1.624 & 13.071 & 2.357 & 12.34 & 4.11$^*$ & 0.598 & 0.542 \\
4658693507772922624 & 80.4737 & -68.6163 & 0.4533 & 1.496 & 12.385 & 2.307 & 10.21 & -1.14 & 0.596 & 0.554 \\
4658854002110866688 & 80.1928 & -67.5869 & 0.7998 & 1.436 & 12.488 & 2.348 & 12.79 & 3.71 & 0.591 & 0.550 \\
4658854831036783232 & 80.1141 & -67.5098 & 0.8646 & 1.472 & 12.744 & 2.327 & 10.52 & -11.26$^*$ & 0.590 & 0.084 \\
4658286933330473600 & 79.8546 & -68.9528 & 0.6250 & 1.621 & 12.260 & 2.324 & 14.00 &   & 0.588 & \\
4658198766214889856 & 79.0002 & -69.1475 & 0.8248 & 1.663 & 12.347 & 2.382 & 15.76 &   & 0.578 & \\
4658293083726796928 & 79.2046 & -69.0403 & 0.7059 & 1.731 & 12.339 & 2.344 & 16.96 &   & 0.574 & \\
4658191172772162944 & 79.5185 & -69.3746 & 1.0326 & 1.596 & 12.702 & 2.394 & 14.90 &   & 0.572 & \\
4658779342721084160 & 78.9703 & -67.8583 & 0.5211 & 1.641 & 12.620 & 2.266 & 17.35 &   & 0.562 & \\
4658806005880090368 & 78.9761 & -67.6454 & 0.7227 & 1.817 & 12.884 & 2.433 & 15.14 &   & 0.553 & \\
4658336548806471936 & 79.4965 & -68.5243 & 0.1823 & 1.578 & 12.169 & 2.310 & 10.41 & -0.96$^*$ & 0.551 & 0.518 \\
4658812053194570240 & 79.1862 & -67.5435 & 0.8066 & 1.636 & 12.858 & 2.237 & 8.80 &   & 0.528 & \\
4658709553769767424 & 80.3100 & -68.1705 & 0.3488 & 1.666 & 12.974 & 2.329 & 17.56 &   & 0.528 & \\
4658275659025982592 & 79.8104 & -69.2550 & 0.9203 & 1.367 & 12.346 & 2.364 & 14.52 &   & 0.515 & \\
4658377810497066368 & 78.8767 & -68.5252 & 0.2905 & 1.717 & 12.960 & 2.257 & 17.61 &   & 0.500 & \\
4658337820116699648 & 79.3460 & -68.4719 & 0.1403 & 1.348 & 12.174 & 2.299 & 9.13 &   & 0.491 & \\
4658743608590786432 & 80.6294 & -67.8497 & 0.6504 & 1.407 & 12.880 & 2.412 & 12.16 & 1.82$^*$ & 0.483 & 0.481 \\
4658785493114963328 & 78.3230 & -67.9287 & 0.5997 & 1.871 & 12.932 & 2.328 & 10.51 & 1.53 & 0.478 & 0.477 \\
4658730070853284608 & 80.7501 & -68.1758 & 0.4955 & 1.269 & 12.647 & 2.367 & 14.24 &   & 0.443 & \\
4658360458826656128 & 78.3023 & -68.7202 & 0.5762 & 1.624 & 13.046 & 2.347 & 12.35 & 7.36 & 0.430 & 0.273 \\
4658737423838030080 & 80.7492 & -68.0642 & 0.5439 & 1.417 & 12.714 & 2.252 & 9.66 &   & 0.422 & \\
4658727528232702976 & 80.8333 & -68.2937 & 0.4986 & 1.316 & 12.636 & 2.272 & 11.23 & 0.98$^*$ & 0.418 & 0.417 \\
4658188350951010560 & 78.9303 & -69.3897 & 1.0669 & 1.831 & 12.295 & 2.309 & 15.28 &   & 0.406 & \\
4658515700465243392 & 80.8487 & -68.3891 & 0.5031 & 1.281 & 12.411 & 2.306 & 11.87 & 1.63$^*$ & 0.384 & 0.383 \\
4658380799796398464 & 78.8235 & -68.4930 & 0.2879 & 1.810 & 12.475 & 2.360 & 18.33 &   & 0.380 & \\
4658276728485892224 & 80.0309 & -69.2272 & 0.9066 & 1.584 & 12.259 & 2.277 & 16.15 &   & 0.379 & \\
4658162241856515712 & 78.7408 & -69.5461 & 1.2338 & 1.636 & 12.438 & 2.368 & 15.65 &   & 0.367 & \\
4658540989235266176 & 81.1009 & -68.2522 & 0.6024 & 1.425 & 12.754 & 2.358 & 13.07 &   & 0.353 & \\
4658833321873098368 & 80.9618 & -67.8727 & 0.7221 & 1.375 & 12.874 & 2.411 & 14.04 &   & 0.345 & \\
4658548337881959424 & 81.0902 & -68.1410 & 0.6263 & 1.523 & 12.747 & 2.430 & 11.91 & 0.82$^*$ & 0.325 & 0.325 \\
4658797244146158976 & 78.2421 & -67.6328 & 0.8495 & 1.813 & 12.513 & 2.330 & 17.56 &   & 0.325 & \\
4658316173421984128 & 80.3508 & -68.7060 & 0.4814 & 1.281 & 12.454 & 2.310 & 18.02 &   & 0.319 & \\
4658515631745778944 & 80.8804 & -68.3981 & 0.5156 & 1.180 & 12.752 & 2.380 & 16.06 &   & 0.312 & \\
4658794838964513792 & 78.2221 & -67.7434 & 0.7636 & 2.022 & 12.619 & 2.279 & 9.74 & -3.73$^*$ & 0.308 & 0.225 \\
\hline
\end{tabular}
\normalsize
\end{table*}

\bsp	% typesetting comment
\label{lastpage}
\end{document}